\documentstyle[12pt]{article}
\def\ni{\noindent}

\def\pr{\em Phys. Rev.}
\def\prl{\em Phys. Rev. Lett.}
\def\be{\begin{equation}}
\def\ee{\end{equation}}
\def\ni{\noindent}
\date{}
\begin{document}
\ni
{\large\bf Hall angle in high-T$_c$ cuprates: Anomalous
temperature dependence and anisotropic scattering\\~\\
N. Kumar}\\
Raman Research Institute, Bangalore 560 080, India\\~\\

\baselineskip=24pt
\ni
{\sl\bf Abstract}.  The anisotropy of the scattering rates,
1/$\tau_{tr} \propto T$ and 1/$\tau_H \propto T^2$, implied
effectively by the anomalous temperature dependence of the
normal-state in-plane Hall angle, $\cot \theta \propto T^2$,
observed in the high-T$_c$ layered cuprates is reasoned out to
be a natural consequence of the semiclassical Boltzmann
transport equation with crossed electric ({\bf E}) and magnetic
({\bf H}) fields. It
is argued that while the scattering rate 1/$\tau_{tr}$ describes
the longitudinal relaxation of the dipolar E-perturbations to the
circular zero-field reference distribution function which is known to
correspond to a non-Fermi liquid with 1/$\tau_{tr} \propto T$,
the scattering rate 1/$\tau_H$ describes the transverse
relaxation of the H-perturbations to the E-induced shifted
reference distribution which is Fermi-liquid-like giving
1/$\tau_H \propto T^2$.  Incorporation of impurity scattering gives
$\cot \theta _H = aT^2 + b$ in agreement with the observed
temperature dependence.\\
$\_\_\_\_\_\_\_\_\_\_\_\_\_\_\_\_\_\_\_\_\_\_\_\_\_\_\_\_\_\_\_\_\_\_\_\_\_\_\_\_\_\_\_\_\_\_$

\ni
The temperature dependence of the normal-state in-plane Hall
angle $\theta_H(T)$ observed in the layered cuprate
superconductors is known$^{1-7}$ to be anomalous in that \(\cot
\theta_H(T) = a + b T^2\)  while the in-plane
resistivity $\rho_{ab}(T)$ is T-linear and the Hall coefficient
$R_H \propto 1/T$.  This is in clear violation of the Kohler scaling
rule,$^{6,7}$ known to be valid generally for conventional
metals, that assumes isotropic scattering rates, i.e., 
$\tau_{tr} = \tau _H$, defined with respect to the 
Hall geometry. Here $\tau _{tr}$ is the usual longitudinal
relaxation time for transport parallel to the applied in-plane
electric field ({\bf E}) while $\tau_H$ is the transverse relaxation
time for transport perpendicular to {\bf E} and the out-of-plane
magnetic field ({\bf H}). Further, it has been found that (i) the anomaly
is generic to  high-T$_c$ layered cuprates, (ii) it is
pronounced for optimally doped curpates with T-linear
$\rho_{ab}(T)$, and (iii) it gets weaker for over doped samples.
Also, no such anomaly is seen in  the case of an  out-of-plane
Hall current.$^3$

Semi-phenomenological approaches to resolving the Hall-angle
puzzle have invoked anisotropic scattering rate, $\tau_H \neq
\tau_{tr}$, following the proposal due originally to
Anderson$^8$ that these two distinct scattering rates arise
naturally from the spin-charge separation of carriers in the
normal state of the layered cuprates. In this note we have
argued out, in generality, the observed 
anisotropy, \(\tau_H \neq \tau_{tr}\), basing on a
re-interpretation of the structure of the Boltzmann transport
equation for crossed {\bf E}- and {\bf H}-fields without any microscopic
particularity.  The basic idea is that while the longitudinal
response to an electric field {\bf E} involves relaxation of a
dipolar diformation  to the circular reference  phase-space
distribution, given to be a non-Fermi liquid with T-linear
\(\rho_{ab}(T)\) implying \(\tau_{tr}^{-1} \propto T\), the
incremental transverse (Hall) response to the add-on cross field
({\bf H}) involves relaxation to a reference state which is now a 
rotated dipolar sub-distribution.  The latter may be viewed as
a dilute Fermi-system {\em per se}, but one with an {\em
excluded} circular phase-space, that makes it a normal-Fermi
liquid with the usual relaxation rate \(\tau _H^{-1} \propto
T^2\). 

Our argument can be appreciated best by reference to the
structure of the Boltzmann equation for the velocity
distribution function under the perturbing crossed {\bf E}- and
{\bf H}-fields. As is well known, the Hall effect is a multiplicative
response where the {\bf E}-field produces a dipolar deformation of the
unperturbed circular distribution function $f_{o,o}({\bf v})$, and
the {\bf H}-field response consists in the rotation of  this
dipolar deformation by an angle, the Hall angle $\theta_H$.  

Let $f_{o,o}({\bf v})$ be the distribution function for {\bf E}
= 0 = {\bf H} and \(\delta f_{E,o}({\bf v})\) be the dipolar
deformation generated by the in-plane electric field {\bf E}.
Then (in obvious notation$^6$)
\be
\delta f_{E,o} ({\bf v}) = - \tau_{tr} e {\bf E\,.\,v} \frac{\partial
f_{o,o}}{\partial\epsilon} 
\ee
Here $\tau_{tr}^{-1}$ denotes the relaxation rate of deformed
distribution $f_{E,o}({\bf v})$ towards the reference circular
distribution $f_{o,o}({\bf v})$. Now, consider the effect of the
crossed out-of-plane-field {\bf H}. This field acts on the dipolar
deformation \(\delta f_{E,o}({\bf v})\) and generates a
rotationally re-deformed distribution \(\delta f_{E,H}({\bf v})\).
(The action of {\bf H} on the original circular reference
distribution \(f_{o,o}({\bf v})\)  generates
(quantum-mechanically) only a diamagnetic response which is not
relevant here.) The question now is how \(\delta f_{
E,H}({\bf v})\) is to relax. The first point to note is that \(\delta f_{
E,H}({\bf v})\) has to relax to the dipolar reference distribution 
\(\delta f_{E,o}({\bf v})\). This reference state characterized
by the dipolar deformation \(\delta f_{E,o}({\bf v})\) may now
be treated as a fermionic sub-system. The latter is certainly
dilute (for low-enough E-field).  Also, the rotational relaxation 
\(\delta f_{E,H}({\bf v}) \rightarrow \delta f_{E,o}({\bf
v})\) is not affected by the background \(f_{o,o}({\bf v})\) which
merely provides an {\em excluded phase-space}. Thus, the relaxation 
\(\delta f_{E,H}({\bf v}) \rightarrow \delta f_{E,o}({\bf
v})\) is essentially characteristic of a normal Fermi liquid. We
must, therefore, have (up to orders linear in {\bf E} \& {\bf H})
\be
\delta f_{E,H}({\bf v}) = \delta f_{E,o}({\bf v}) + H\tau_H \frac{e}{c}
{\bf v} \times \hat{H} \frac{\partial v}{\hbar \partial
{\bf k}}\,\,\frac{\partial}{\partial {\bf v}}\,\,
\delta f_{E,o}({\bf v})
\ee
with $\hat{H}$ a unit vector along {\bf H}, and \(\tau_H^{-1}
\propto T^2\) as for a normal Fermi liquid without impurity scattering.

\ni
From Eqs. (1) and (2), we at once have for the Hall angle
$\theta_H$: 
\be
\cot \theta_H \propto \frac{1}{\tau _H} 
\ee
Now, for an impure normal Fermi-liquid \(\tau_H^{-1} = \alpha T^2 +
\beta\), where $\beta$ denotes the effect of impurity scattering.
Hence, 
\be
\cot \theta_H  =  aT^2 + b
\ee
as indeed observed by Chien {\em et al.}$^1$ in Zn-doped YBCO
single crystals, and subsequently reported more widely.$^{2-7}$

The above argument is consistent with the absence of anomaly
when the Hall current is out-of-plane (i.e., along the C-axis)
with the crossed-fields in-plane.  This is because the C-axis
transport has been shown to be controlled intrinsically by the
in-plane transport in the normal state, and hence there is a
single relaxation time \(\tau_{tr} \propto T^{-1}\) consistent
with the T-linear in-plane resistivity.$^{9-11}$

We would like to conclude with the following comments.
The main point of our argument is that while the system may be a
strongly correlated one with non-Fermi liquid characteristics,
e.g., the T-linear in-plane resistivity, a small deformation (in
the sense of distribution function) of the system when probed
appropriately may behave differently, and in particular as a
normal Fermi-liquid. This notion is, of course, somewhat
familiar in terms of the idea of the electron- or the
hole-pockets of a complex Fermi-surface representing sub-sets of
carriers with different characteristics, e.g., effective masses,
etc.  In our case 
of the Hall angle, the electric field prepares the small
deformation (sub-set of carriers) and the magnetic-field probes
it.  It should be 
possible to extend this argument to other, possibly multipolar,
deformations.

\end{document}